\documentclass[proceedings]{rmaa}

\usepackage{paralist}

\usepackage{psfrag,color}

\SetYear{2008} \SetConfTitle{A Long Walk Through Astronomy:  A
Celebration of Luis Carrasco's 60th Birthday}

\title{The energetic environment and the dense interstellar
medium in ULIRGs}

\author{
  O. Vega,\altaffilmark{1,2}
  M.S. Clemens,\altaffilmark{2} A. Bressan,\altaffilmark{1,2,3} G.L.
  Granato,\altaffilmark{2,4} L. Silva,\altaffilmark{4}
  and P. Panuzzo\altaffilmark{2,5}}

\altaffiltext{1}{INAOE, Luis Enrique Erro 1, 72840 Tonantzintla,
Puebla, Mexico (ovega@inaoep.mx).}

\altaffiltext{2}{INAF, Osservatorio Astronomico di Padova, Vicolo
dell'Osservatorio, 5, 35122 Padova, Italy}

\altaffiltext{3}{SISSA, Strada Costiera, I-34131 Trieste, Italy}
\altaffiltext{4}{INAF, Osservatorio Astronomico di Trieste, Via
Tiepolo 11, I-34131 Trieste, Italy}

\altaffiltext{5}{Laboratoire AIM, CEA/DSM - CNRS - Universit\'{e}
Paris Diderot, DAPNIA/Service d'Astrophysique, B\^{a}t. 709,
CEA-Saclay, F-91191 Gif-sur-Yvette C\'{e}dex, France}
\shortauthor{Vega, Clemens, Bressan, et al.}

\shorttitle{The energetic environment of ULIRGs}

\listofauthors{O. Vega, M. S. Clemens, A. Bressan, G.L. Granato, L.
Silva \& P. Panuzzo}
\indexauthor{Vega, O.} \indexauthor{Clemens, M. S.}
\indexauthor{Bressan, A.} \indexauthor{Granato, G. L.}
\indexauthor{Silva, L.} \indexauthor{Panuzzo, P.}

\abstract{We fit the near-infrared to radio spectral energy
distributions of a sample of 30 luminous and ultra-luminous infrared
galaxies with  models that include both starburst and AGN
components. The aim of the work was to determine important physical
parameters for this kind of objects such as the optical depth
towards the luminosity source, the star formation rate, the star
formation efficiency and the AGN fraction. We found that although
about half of our sample have best-fit models that include an AGN
component, only 30\% have an AGN which accounts for more than 10\%
of the infrared luminosity whereas all have an energetically
dominant starburst. Our models also determine the mass of dense
molecular gas. Assuming that this mass is that traced by the HCN
molecule, we reproduce the observed linear relation between HCN
luminosity and infrared luminosity found by Gao \& Solomon (2004).
However, our derived conversion factor between HCN luminosity and
the mass of dense molecular gas is a factor of 2 smaller than that
assumed by these authors. Finally, we find that the star formation
efficiency falls as the starburst ages.}

\resumen{Presentamos los ajustes a la distribuci\'on espectral de
energ\'{\i}a, desde el cercano-infrarrojo hasta radio, de 30
galaxias luminosas y ultraluminosas en el infrarrojo con modelos que
incluyen  formaci\'on estelar (SB) y actividad nuclear (AGN). La
motivaci\'on de este trabajo fue determinar importantes par\'ametros
f\'{\i}sicos para esta clase de objetos, como son: la profundidad
\'optica hacia la fuente infrarroja, el ritmo y la eficiencia de la
formaci\'on estelar y  la contribuci\'on del AGN. Encontramos que,
aunque la mitad de la muestra necesita modelos combinados (i.e.
AGN+SB), solo el 30\% necesit\'o un AGN contribuyendo m\'as del 10\%
a la luminosidad infrarroja. En todos los casos  la contribuci\'on
del SB es dominante. De nuestros modelos tambi\'en pudimos obtener
 la masa de gas molecular denso. Si asumimos  que
esta masa es la misma  que la que es trazada por la mol\'ecula de
HCN, podemos reproducir la relaci\'on lineal entre la luminosidad de
HCN y la masa de gas denso observada por Gao y Solomon (2004). Sin
embargo, el factor de conversi\'on entre la luminosidad de HCN y la
masa de gas denso que derivamos es un factor 2 menor que el asumido
por estos autores. Por \'ultimo, tambi\'en  encontramos que la
eficiencia de la formaci\'on estelar disminuye a medida que el brote
de formaci\'on estelar envejece.}

\addkeyword{Ulirgs} \addkeyword{ISM: dust extinction}
\addkeyword{Galaxies: active} \addkeyword{Infrared: galaxies}

\begin{document}
\maketitle

\section{Introduction}
\label{sec:intro}

With total infrared luminosities between $10^{11}- 10^{12}$
L$_\odot$ and $\geq 10^{12}$ L$_\odot$, respectively, Luminous and
Ultraluminous Infrared galaxies, (U)LIRGs, are the most luminous
objects in the local universe. Although scarce at low redshift, they
may account for the bulk of all star formation activity at $z > 2 -
3$ and dominate the far-infrared background (e.g. Blain et al.
2002). Many of them are found in merging systems (e.g. Sanders et
al. 1988), suggesting  that dynamical interaction has driven gas
towards the nucleus, fueling a massive starburst (SB) and/or the
central massive black hole (Mihos \& Hernquist 1996). Despite
extensive investigation over the last decades, there is still
considerable uncertainty as to the nature of their power source. In
order to analyze the power mechanism in these sources, tracers that
do not suffer large extinction have been preferred recently, such as
hard X-rays, and MIR to radio diagnostics. The weakness of the hard
X-ray luminosity seen in most ULIRGs have been interpreted  by
Risaliti et al. (2006) as a real lack of AGN activity. MIR
diagnostics such as the strengths of PAH emissions, the 9.7${\mu}$m
absorption feature, the high ionization lines, and the MIR continuum
slope indicate that the starburst is dominant in about 80\% of
ULIRGs. However,  there are strong discrepancies in the results
obtained with the different methods (e.g. Armus et al. 2007). This
is in line with the recent finding that the strength of PAHs and the
shape of the mid-infrared continuum cannot be safely used to
disentangle AGN and SB contributions (Vega et al. 2005, Weedman et
al. 2005). In the radio domain,  the most direct way of
distinguishing between AGN and starburst power sources in ULIRGs is
to search for very compact radio continuum emission towards the
nuclei. Nagar et al. (2003), by using 15~GHz radio continuum data
with a resolution of 150 mas, concluded that most of the 83 ULIRGs
of their sample are AGN powered. However, Smith et al. (1998)
detected radio-SNe in the NW nucleus of Arp~220 in a region of $0\,
\farcs 2 \times 0\, \farcs 4$ and concluded that no AGN is necessary
to explain the IR luminosity in this source.

We followed a different approach to examine the energetic
environment in (U)LIRGs. Instead of considering a single spectral
region, we base our study on the thorough analysis of the
panchromatic spectral energy distribution (SED). Thus, we first
determine the observed NIR - radio SED of a sample of 30 local
(U)LIRGs and then, these data are compared with suitable models with
SB and AGN components.

A Similar approach has been adopted by Farrah et al. (2003). They
used SB and AGN templates to model the SED from the optical-submm
range. We prefer to use models built on well calibrated star forming
complexes instead of templates re-scaled to different luminosities
because it avoids the dubious process of "re-scaling" a "template"
to another luminosity, and  because the use of models gives access
to a more sound physical picture of the environment within which
such objects evolve, that can be used for the study of even more
extreme phases such as those likely found at high redshift
(e.g. Mortier et al. 2005), which is
one of the ultimate goals of the present investigation. We excluded
from the analysis the optical data because, due to the large
extinction, they are likely dominated by the underlying old stellar
population and not by the starburst (e.g. Farrah et al. 2001).
In contrast, the inclusion of radio data in the analysis gives us an independent
constraint on the SB strength and age (Bressan, Silva \& Granato
2002).
\begin{table}[t!]
\centering
 \caption{GRASIL parameters for the SB SED library} \label{tab:chevo}
\setlength{\tabnotewidth}{\columnwidth}
  \tablecols{6}
  \label{tab:gra}

\begin{tabular}{cccccccc} \hline
\hline

$t_{\rm{b}}$ &
age$_{\rm{b}}$& $\beta$&$t_{\rm{esc}}$ & $\tau_{1}$&$f_{\rm{mol}}$ &r\\
 Myr&Myr    &&Myr&
    &&kpc\\
  \hline
5-80 &1-315& 1-2&5-80&9-180&.1-.9 &0.01-1\\

\hline
\end{tabular}
\end{table}
\begin{figure*}[t!]
\centering
  \includegraphics[width=0.99\columnwidth]{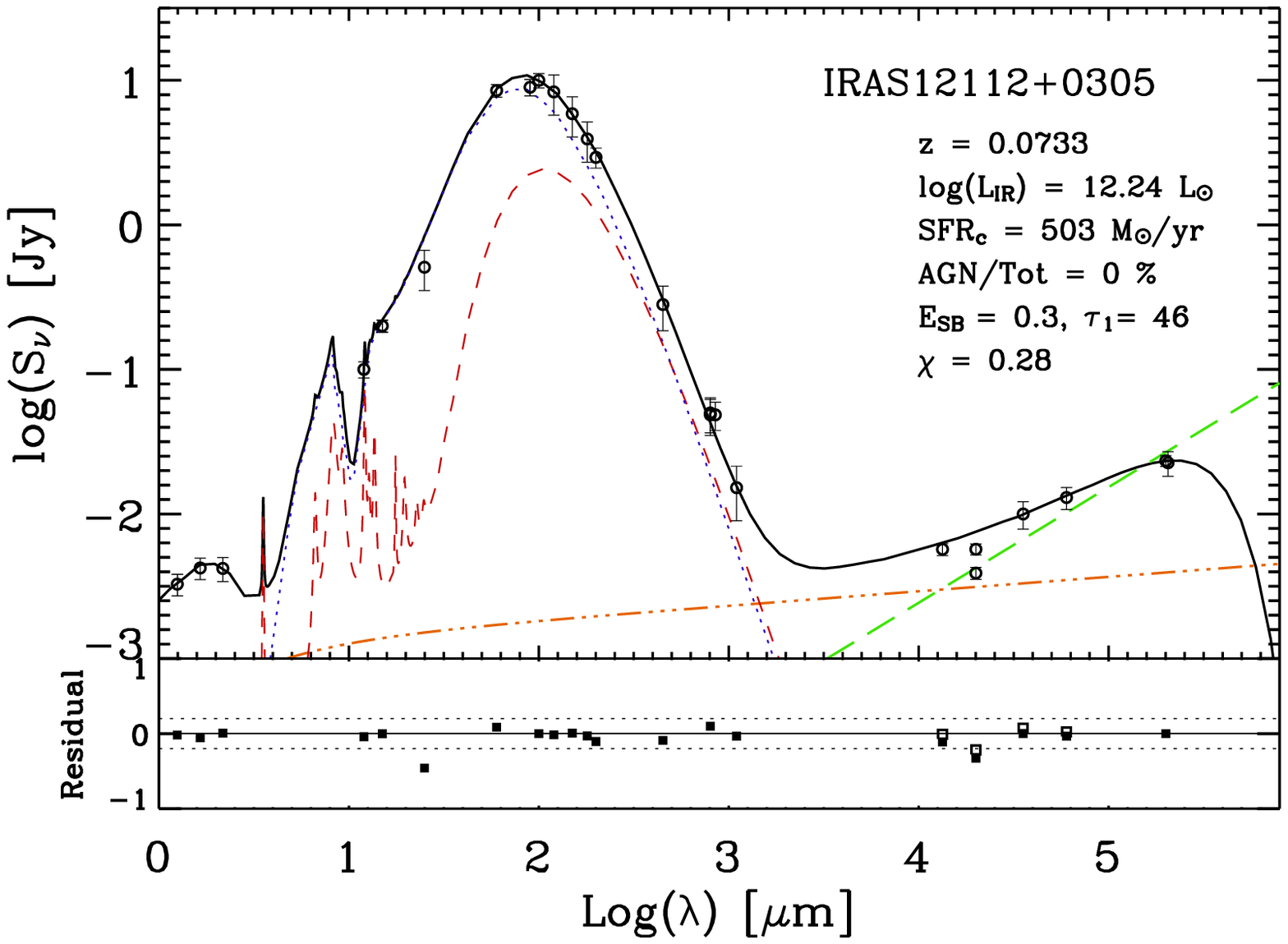}
 \includegraphics[width=0.99\columnwidth]{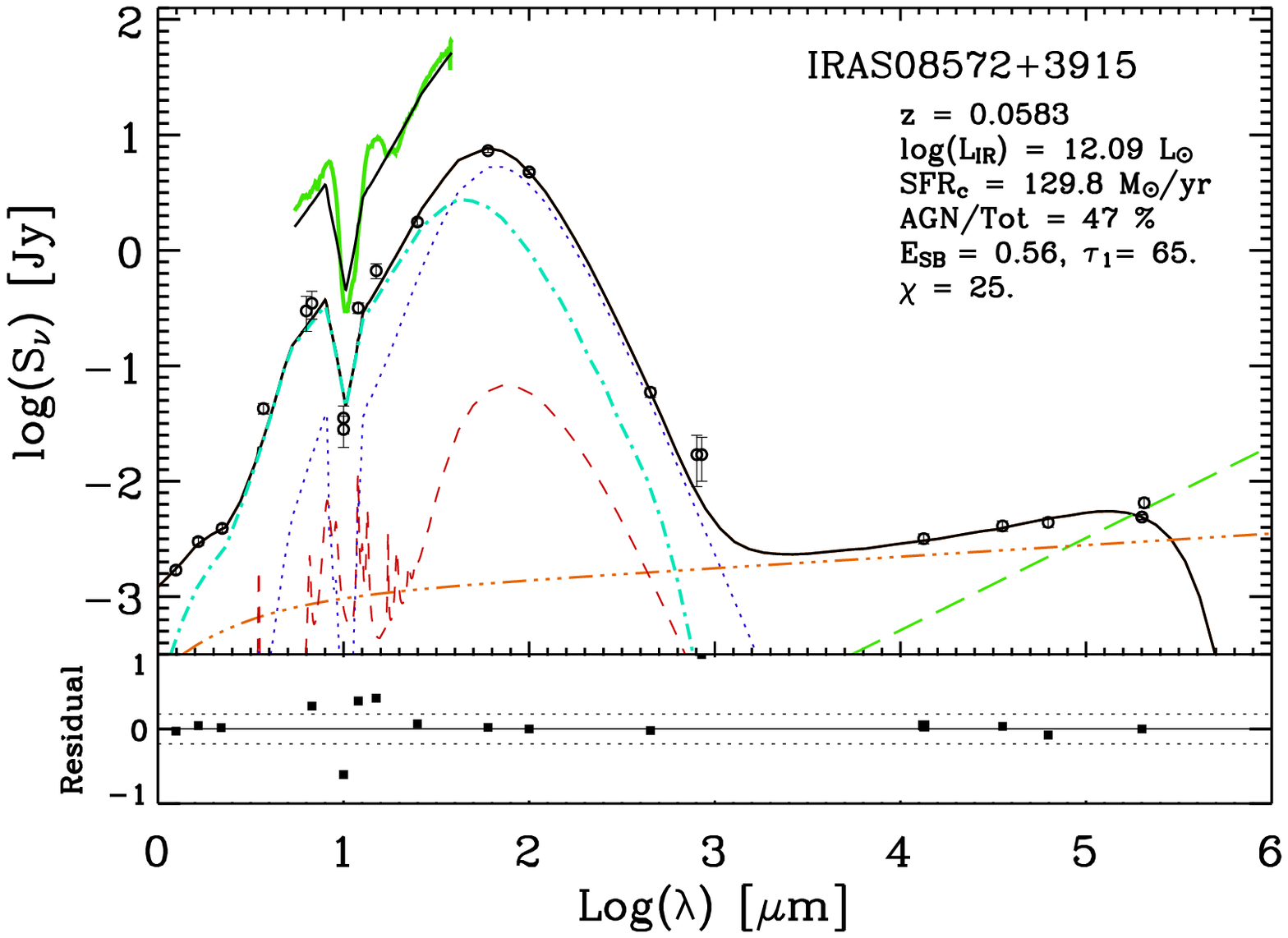}
  \includegraphics[width=0.99\columnwidth]{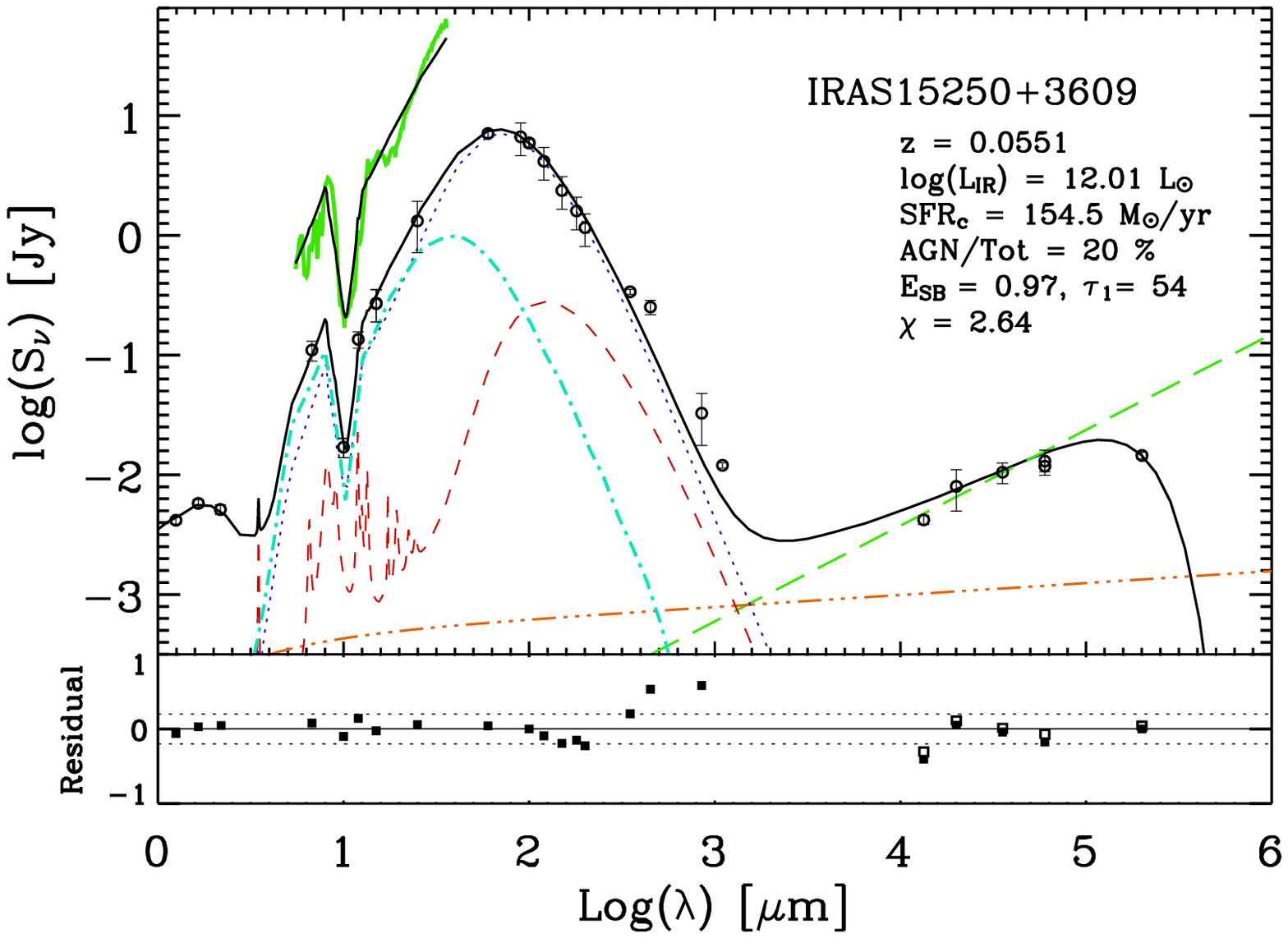}
   \includegraphics[width=0.99\columnwidth]{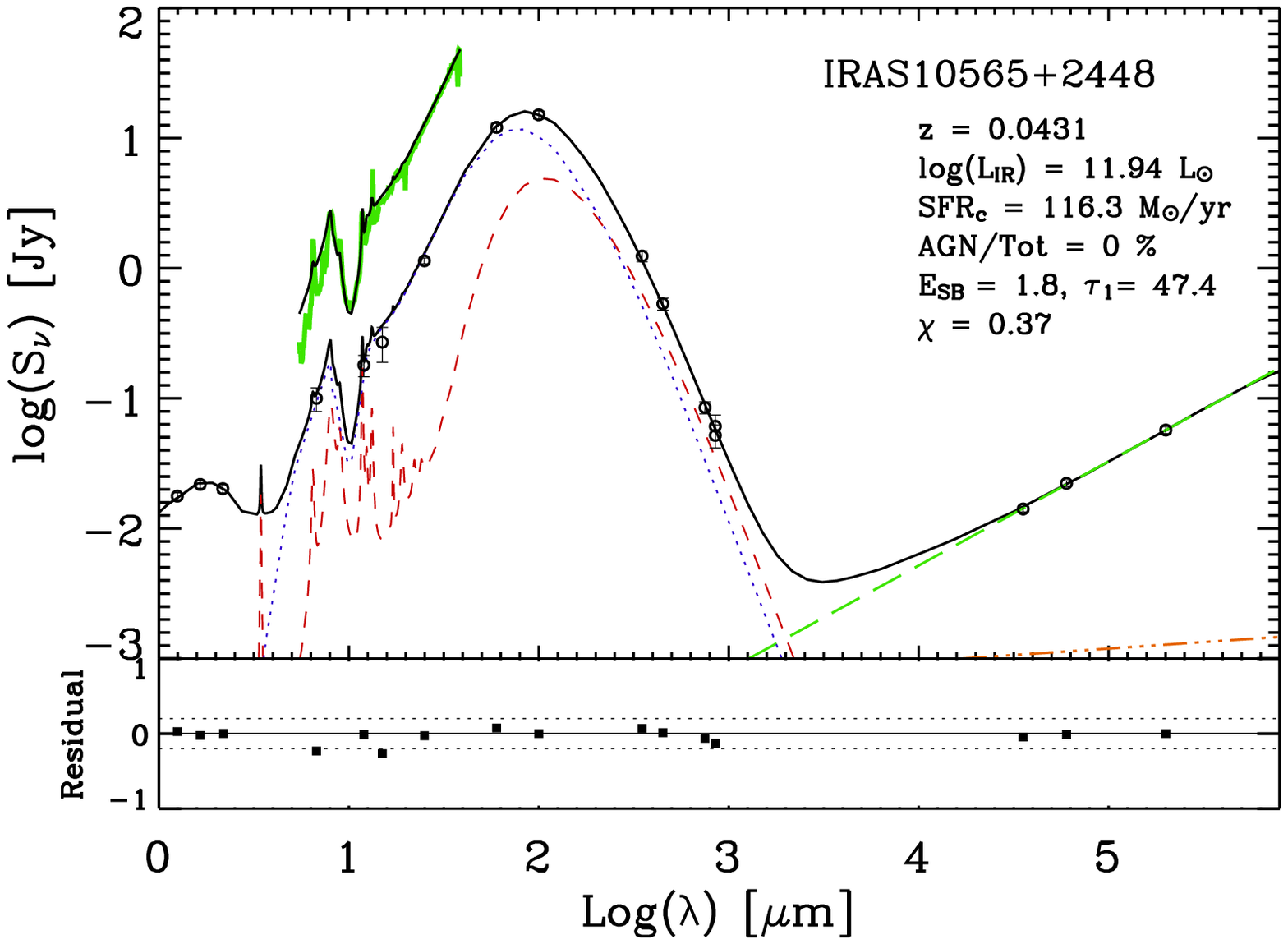}
    \includegraphics[width=0.99\columnwidth]{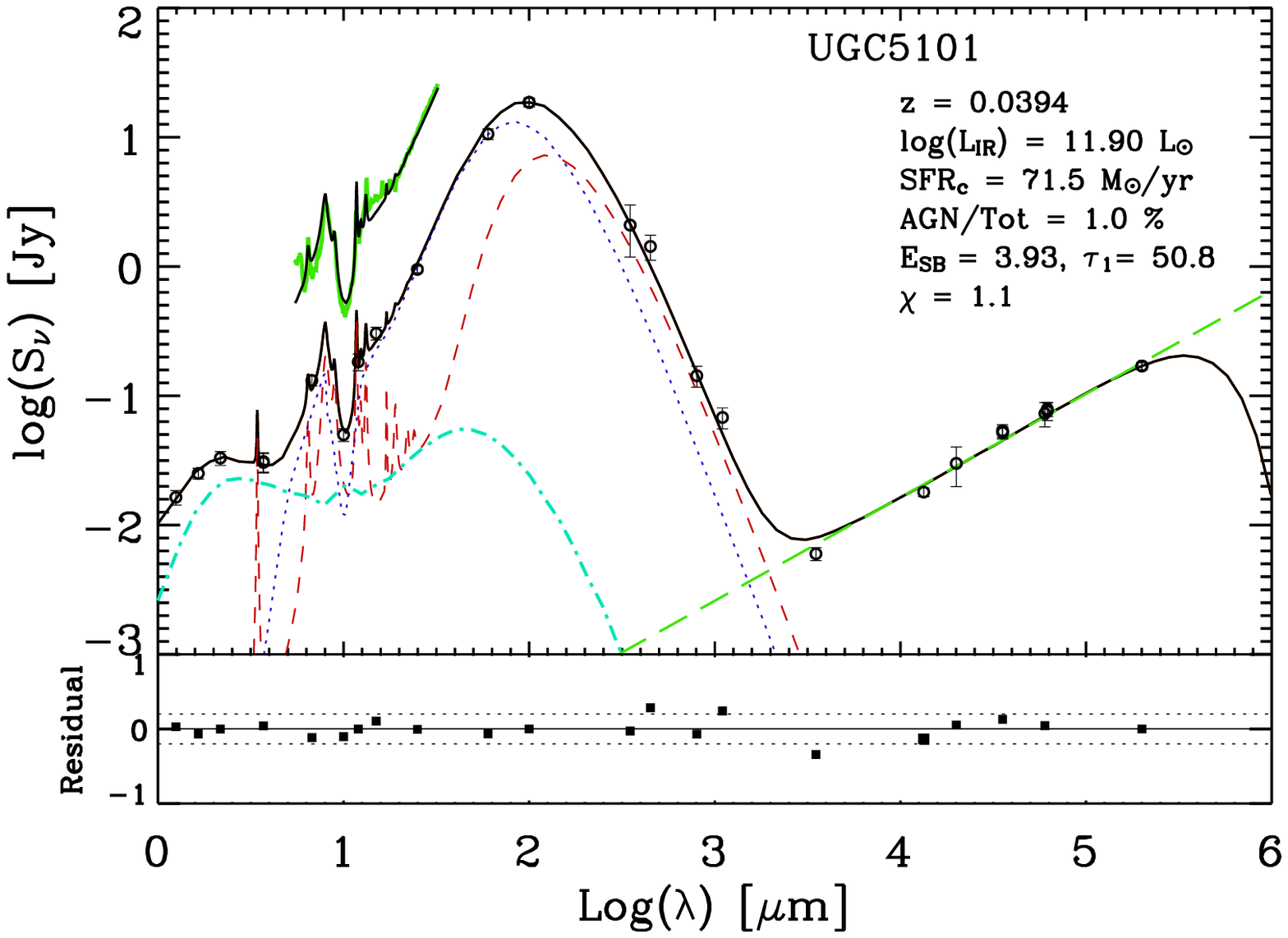}
  \includegraphics[width=0.99\columnwidth]{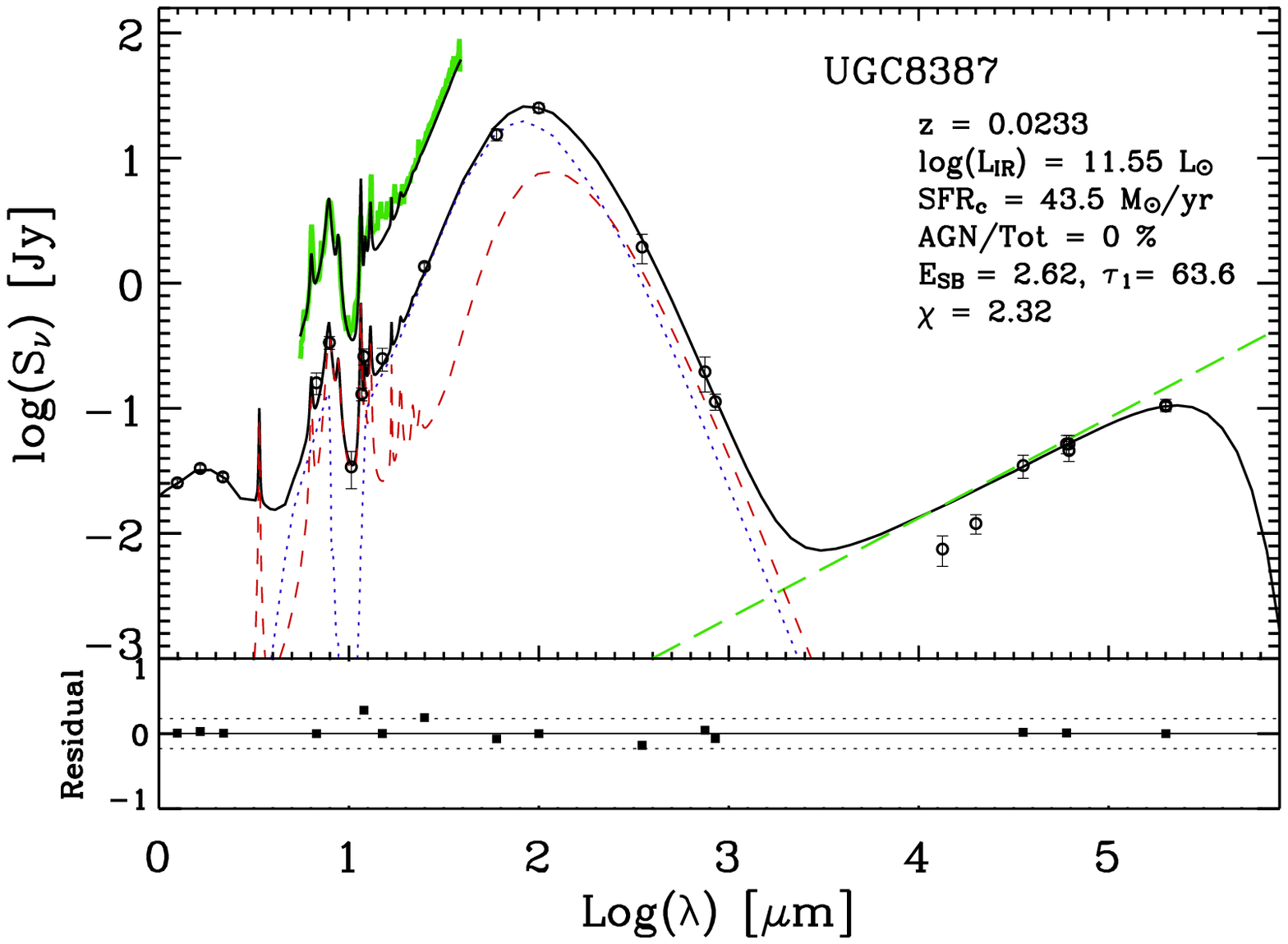}
  \caption{Examples of the comparison between the broad band SED (open circles)
 and the best fit model (thick solid line) for six galaxies of our sample. The additional emission from the
AGN, when needed, is indicated by a dot-dashed line. When available,
the IRS low resolution spectrum (solid thick line) and the
corresponding SED model (solid thin line) are shown upward displaced
by one dex. The residuals between models and data are shown in the
bottom panels as $(f_{\rm{data}}-f_{\rm{model}})/f_{\rm{data}}$. }
  \label{fig:sb_sed}
\end{figure*}
\section{Sample}
\label{sec:sample}

 The selection criteria for our sample of 30 (U)LIRGs  are
summarized as follows: (a) The galaxies are compact (U)LIRGs from
Condon et al. (1991) (i.e. brighter than $5.25\;\rm Jy$ at $60\;\rm
\mu m$) with  radio data at 3 or more frequencies. (b) We also
require the galaxies to have a well-sampled infrared SED, with data
in the NIR, MIR and FIR spectral ranges. For all the sources of our
sample we collected NIR data from the 2MASS All-Sky Extended Source
Catalog and IRAS data from IRSA and NED, respectively. Radio data
were taken from Clemens et al. (2008). Additional IR and sub-mm
broad-band data were taken from the literature. We also found
\emph{Spitzer} IRS spectra in the SSC archive for eleven objects of
our sample. Our objects are compact and isolated so that we are
confident that our fluxes sample the entire starburst in all cases.

\section{Models}
\label{sec:command} In order to find the best fit model we adopt the
following approach:  we first test the null-hypothesis that our
sample sources are pure SB. Then, if the observed SED shows an
excess emission over the best fit in those wavelength regions that
are more affected by the  presence of an embedded AGN, typically in
the NIR-FIR range, we reject the null-hypothesis and add an AGN
component to the model. The AGN model is selected from a library
generated with the radiative transfer code developed by Granato \&
Danese (1994) and added to the SB model in such a way that the total
luminosity is
\begin{equation}
L_{\lambda}^{\rm{tot}}= (1-f)\times L_{\lambda}^{\rm{SB}}+f \times
L_{\lambda}^{\rm{AGN}} \  {\rm{ , with }} \ 0\leq f \leq 1.
\end{equation}
Notice that in this work our aim is just to assess the possible
contribution of the warm AGN component to the total SED of ULIRGs,
without any pretense of discriminating the precise AGN properties.
The best fit is obtained by minimizing the merit function, $\chi$,
calculated as
\begin{equation}\label{equ:chi}
\chi
=\frac{1}{N}\sum_{i=1}^{N}\left(\frac{F_{mod}(i)-F_{obs}(i)}{Err(i)}\right)^2,
\end{equation}
where $F_{mod}(i)$, $F_{obs}(i)$, and $Err(i)$ are the model flux
values,  the observed fluxes and observational errors respectively.
$N$ is the number of passbands used for the fit.

The SB models  are selected from a library generated with
GRASIL\footnote{GRASIL can be run via the WEB interface
web.pd.astro.it/galsynth/ which is maintained at OAPD, INAF by L.
Paoletti, A. Petrella \& D. Selvestrel, and can also be downloaded
from the webpage http://adlibitum.oats.inaf.it/silva/default.html}
(Silva et al. 1998) and involve seven main free parameters: the
e-folding time and age of the burst, $t_{\rm{b}}$ and
$age_{\rm{b}}$, the sub-mm dust spectral index, $\beta$, the escape
time, $t_{\rm{esc}}$, the molecular cloud (MC) optical depth at 1
$\mu$m, $\tau_{1}$, the fraction of gas in MCs, $f_{\rm{mc}}$, and
the core radius of the King profile for the dust-star distributions,
$r$. The range of values of these free parameters are listed  in
Table \ref{tab:gra}. For a complete description of models we refer
the reader to Vega et al. (2008) and references therein.

\section{Results and Conlusions} \label{sec:results}

\begin{figure}[h]
  \includegraphics[width=1.\columnwidth]{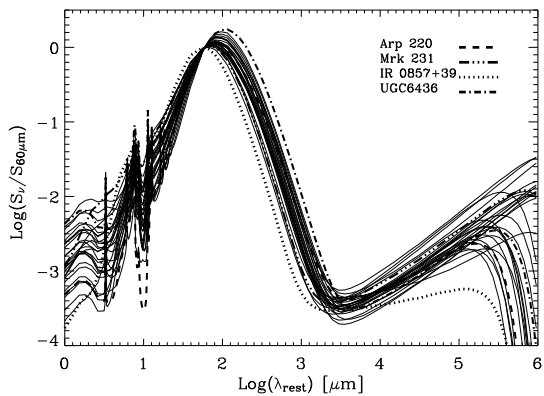}
  \caption{Comparison of SED models for our sample of (U)LIRGs normalized to the 60 $\mu$m flux density. }
  \label{fig:sed60}
\end{figure}

\begin{table*}[!t]
\setlength{\tabnotewidth}{2.1\columnwidth}
  \tablecols{12}
\small \caption{Best fit parameters for the starburst
model\tabnotemark{a}. } \label{tab:fitpara}
\begin{tabular}{lccccclc|lcc|c} \hline \hline
NAME
&$t_{\rm{b}}$&log($age_{\rm{b}}$)&$\beta$&$t_{\rm{esc}}$&$\tau_{\rm{1}}$&
$f_{mol}$&    r& $E_{\rm{SB}}$&SFR&$M_{\rm{den}}$&L$_{\rm{AGN}}$/L$_{\rm{IR}}$\\
&(Myr)&(yr)&&(Myr)&&&(kpc)&&($\frac{M_\odot}{yr}$)&(M$_\odot$)&$\%$\\
 \hline
Mrk~231$^*$&35&7.572&2.00&26&33.13&0.65&0.40&1.07&436&10.1&16\\
IR~14348-1447&10&7.170&1.95&14&49.09&0.26&0.27&1.48&336&9.9&8\\
IR~12112+0305&    25&6.880&2.00&7& 45.80&     0.35&      0.20&0.30&503&10.0&--\\
IR~05189-2524$^*$&10&6.750&2.00&6&27.37&0.40&0.65&0.56&226&9.6&17\\
Arp~220$^*$&           35&7.340&1.80&20&64.93&     0.14&      0.55&0.62&235&9.8&--\\
IR~08572+3915$^*$& 7&6.590&2.00&3&64.93&0.95&0.25&0.56&130&9.1&$47$\\
UGC~8696$^*$& 20&7.650&1.95&37&56.72&0.20&0.37&2.23&138&9.7&$11$\\
IR~15250+3609$^*$&    20&7.289&2.00&16& 53.66&     0.47&      0.70&0.97&154&9.4&$20$\\
IR~10565+2448$^*$&  45&7.910&2.00&60&      47.43&     0.29& 0.14&1.80&116&9.7&--\\
UGC~5101$^*$& 7&7.440&1.95&22&50.85&0.18&0.20&3.93&71&9.8&1\\
IZW~107&      30&7.770&2.00&40&33.13&0.17&0.27&1.96&87&9.5&--\\
IR~01364-1042& 10&7.290&2.00&14&116.1&0.53&0.50&1.95&92&9.4&$11$\\
IR~10173+0828&  40& 7.200&2.00&14&42.22&0.27&0.41&0.41&110&9.4&--\\
Arp~299&25&7.756&2.00&43&31.75& 0.17&  0.70&2.28&67&9.4&--\\
UGC~4881&      40&7.986&2.00& 44&49.97&     0.22&       0.20&2.42&54&9.4&-- \\
CGCG~436-30$^*$&  15&7.500&2.00&24&45.09&0.15&0.60&2.11&58&9.2&$5$\\
IC~1623&      25&7.635&1.80& 24&39.63&     0.17&      0.51&1.73&65&9.4&-- \\
NGC~1614&  20&7.840&2.00&50&33.13& 0.17&0.30&3.46&29&9.1&$18$ \\
UGC~8387$^*$&   20&7.720&        1.95&33&      63.65&     0.28&      0.20&2.62&43&9.5&--\\
NGC~7469&  30&7.938&2.00&45&33.13&0.16&0.43&2.89&41&9.2&$10$\\
UGC~2369 & 25&7.840&2.00& 36&      31.70&     0.12&      0.67&2.77&47&9.4&-- \\
IIIZW~35& 45& 7.545&1.95&35&61.20&     0.18&     0.55&0.78&60&9.3&--\\
IC~5298& 20&7.900&2.00&60&30.88&0.10&0.42&4.00&25&9.3&--\\
Arp~148 & 20&7.735& 1.80&27&75.07&     0.20&      0.24&2.72&37&9.4&--\\
NGC~2623$^*$&35&7.726&2.00& 30&71.94&0.32&  0.35&1.52&47&9.3&$2$ \\
Mrk~331&45&7.910&2.00&35&53.66&0.14&0.70&1.80&38&9.2&$5$\\
NGC~34&  20&7.875&2.00&65&42.23&0.14&0.45&3.75&21&9.1&$4$\\
NGC~5256&15&7.630&1.75&15&45.09&0.30&0.60&2.84&40&9.3&5 \\
UGC~6436&       30& 7.990&1.90&34&40.89&     0.05&     0.37&3.25&25&9.1&--\\
NGC~6286& 10&7.640&1.80&23&      94.25&     0.50& 0.47&4.36&14&9.3&$5$\\
\hline \tabnotetext{a}{Galaxies with an $^*$ have \textit{Spitzer}
spectra (projects IDs 105, and 14, P.I.: J. R. Houck, and ID 30323,
P.I.: L. Armus).}
\end{tabular}

\end{table*}
In Fig. \ref{fig:sb_sed} we show the SED fits for six galaxies of
our sample. The corresponding best-fit parameters of the SB
components for all the sample are shown in Table \ref{tab:fitpara}.
The last four columns of the table  display the value of the
evolutionary phase of the SB, $E_{\rm{SB}}$, the current star
forming rate, SFR, the mass of the dense component, $M_{\rm{den}}$,
and the fractional contribution of the AGN to the 8-1000 $\mu$m flux
indicated as AGN/Tot. The SED fits for all the sample can be seen in
Vega et al. (2008) \footnote{SED models and data are available in
electronic form at http://www.aanda.org, and they can also be
downloaded from
http://www.adlibitum.oats.inaf.it/silva/default.html.}. In Fig
\ref{fig:sed60} we plot all the SED best-fit models of our sample of
(U)LIRGs normalized to the 60 $\mu$m flux density. The plot shows
the large heterogeneity of (U)LIRGs SEDs. In particular, Arp~220,
which is usually used as prototype of ULIRGs, has one of the most
extreme SED.

\subsection{Star formation versus AGN}
The 53\% of the sources show evidence for the presence of an AGN.
However, only in 9/30 sources
 the contribution of the AGN to the total IR luminosity
  exceed 10\%.
  Only one galaxy, IR~08572+3915, seems to be  completely powered by an AGN.
 We therefore find that the fraction of SB dominated objects in our sample (i.e.
 pure SB galaxies and mixed galaxies with AGN contributions  $<10\%$)
  is 70\%. Our measured AGN fractions are consistent with
those reported by Armus et al. (2007) based on the MIR line ratios
Ne[V]/Ne[II] and O[IV]/Ne[II]. But, the PAH equivalent widths and
mid-infrared spectral slope overestimate (often by large factors)
the AGN fraction. Therefore, our work supports the notion that
\emph{neither the mid-infrared slope nor the PAH equivalent widths
provide a good estimate of the AGN contribution} (see conclusions in
Vega et al. 2005). For two galaxies, our findings are markedly
different to those of Armus et al., namely IR~14348-1447 and
IR~15250+3609. For the former, we find an AGN contribution of $\sim
10 \%$, while they  find no evidence of an AGN from the MIR
analysis. Indeed, this AGN contribution is required by its red NIR
colour, J-K$ = 1.71$ which cannot be explained by stellar
populations alone. The power source of IRAS~15250+3609 could not be
determined by Armus et al., while our analysis indicates that it is
a mixed object with a very obscured AGN, that contributes $\sim 20
\%$ of the IR luminosity. This is where the combined analysis of the
IRS spectra and panchromatic SED is most powerful.
\begin{figure}[!h]\centering
  \vspace{0pt}
  \includegraphics[width=\columnwidth]{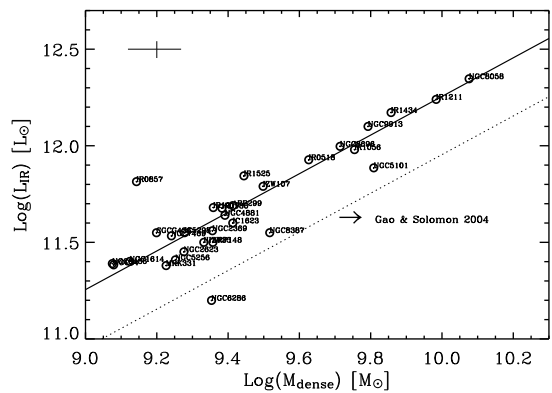}
  \caption{Correlation between the dense gas  mass,
derived from our  models, and the AGN corrected IR luminosity for
the 30 (U)LIRGs.  The thick solid line is the linear fit. The dotted
line corresponds to the fit between the IR luminosity and the dense
molecular mass as traced by HCN emission obtained by GS04. The
cross in the upper left denotes the typical errors of the IR
luminosities and masses derived with our models}
  \label{fig:mol}
\end{figure}
\subsection{Molecular masses}

The MIR-FIR spectral region in our (U)LIRGs is dominated by the
emission from the molecular clouds where the star formation is
taking place. Therefore we can determine quite accurately the mass
in dust associated with the star forming regions via the extinction
required to reproduce the infrared part of the SED. The dust mass
can  be converted into molecular gas mass by using a typical
dust/gas ratio (i.e. G/D =100). We find a relation between the
infrared luminosity and molecular gas mass with the same slope as
that derived by Gao \& Solomon (2004, hereafter GS04) from
observations of HCN emission, see Fig \ref{fig:mol}. However, we
derived a constant of conversion between the HCN luminosity and the
mass of dense molecular gas a factor of 2 smaller than that assumed
by these authors for their global sample of galaxies, but in
agreement with the conversion factor found by Graci\'a-Carpio et al.
(2006) for their (U)LIRG population. Actually, an accurate
determination of the conversion factor between HCN and the mass of
the very dense molecular gas needs more extensive studies, including
the combined analysis of the SED and the high excitation HCN
transitions (Vega et al. 2009 in prep.).
\begin{figure}[!t]
  \includegraphics[width=0.99\columnwidth]{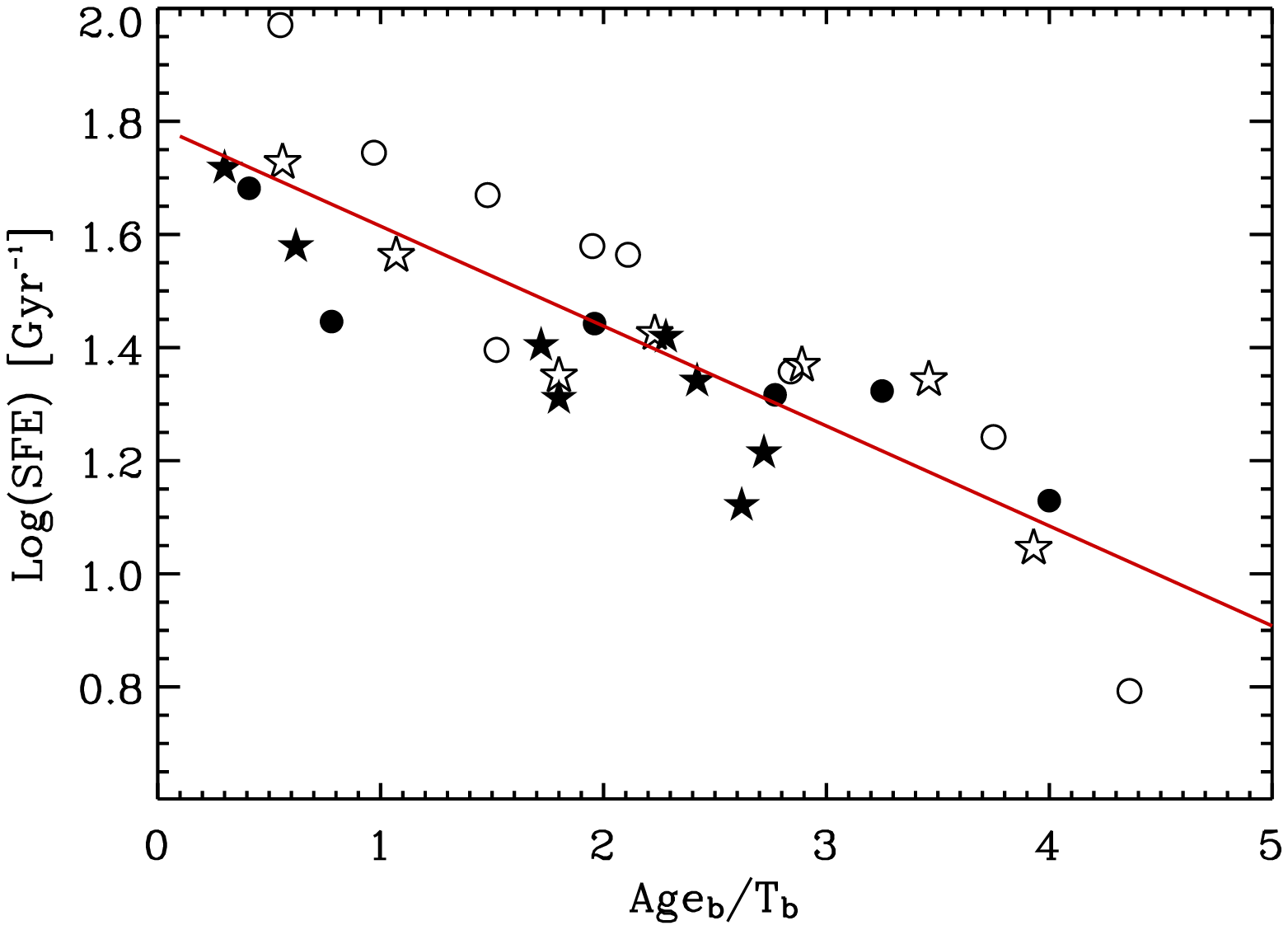}\\
\includegraphics[width=0.99\columnwidth]{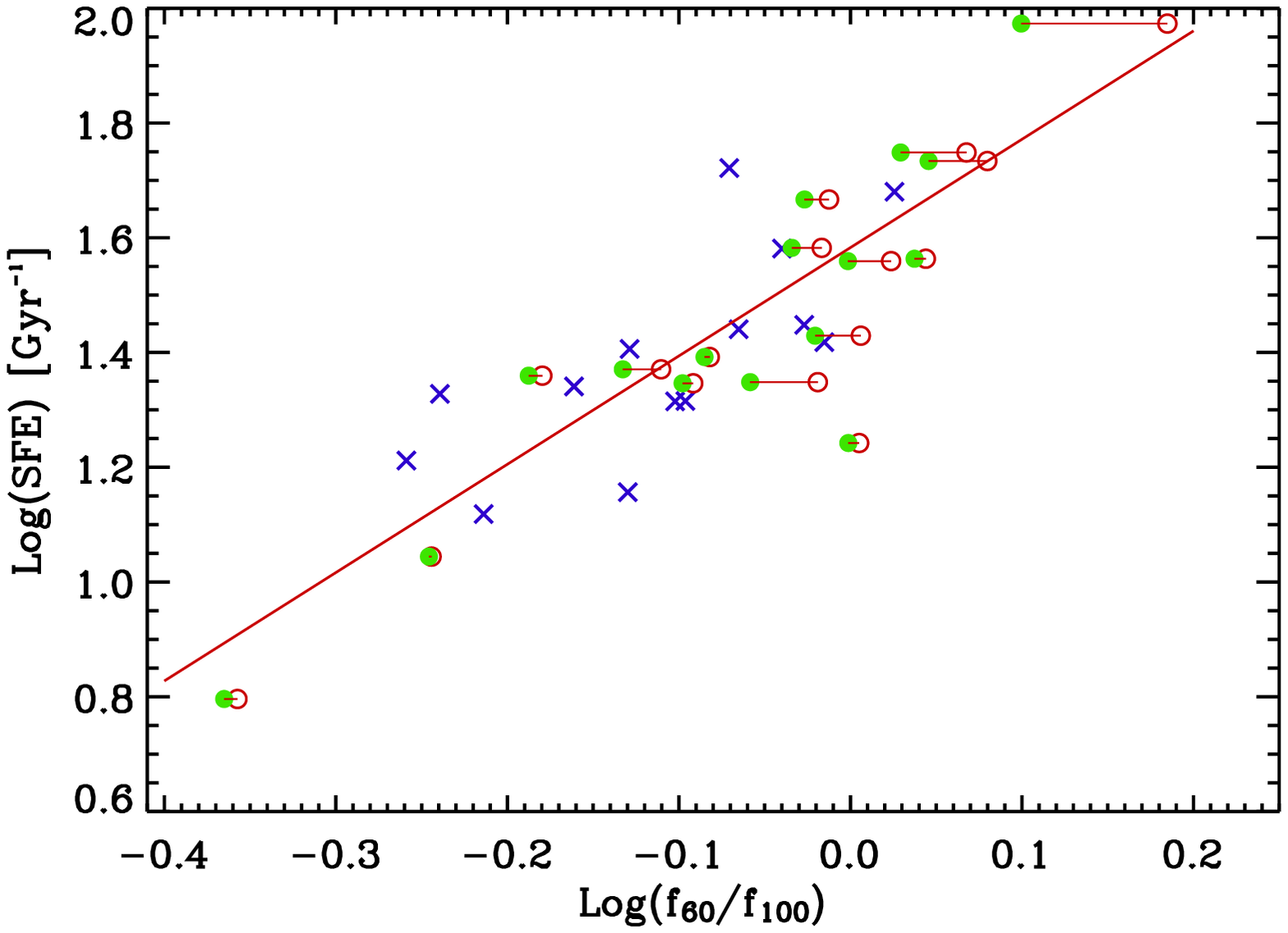}

  \caption{\textit{Upper panel: }SFE as derived from our models versus
the evolutive phase of the starburst. Galaxies in common with the
GS04 sample are marked as stars. Filled symbols correspond to
galaxies fitted with pure SB models while open symbols correspond to
those fitted with mixed models. \textit{Lower panel: }Correlation
between the far-infrared color
 ($f_{60}/f_{100}$) and the derived SFE for our sample of (U)LIRGs. Crosses
denote the pure SB galaxies, while open circles denote the galaxies
with AGN. The filled circles denote the  mixed galaxies but with the
IR colours corrected for the AGN contributions.}
  \label{fig:sfe}
\end{figure}

\subsection{The star formation efficiency}
With the ratio between the SFR and the M$_{\rm{den}}$, we obtain a
quantity that measures the reciprocal of the SF time scale and that
is usually called star formation efficiency (SFE, GS04). We do not find
clear correlations between SFE and IR luminosity
of the SB, or the mass of molecular gas.
However we do find
an evident anti-correlation between SFE
and the SB phase ($E_{\rm{SB}}=\frac{age_{\rm{b}}}{t_{\rm{b}}}$, see
Vega et al. 2005, for a thorough discussion on the SB phases). The
SFE decreases by about one order of magnitude as the SB evolves from
its early phase to the more advanced phases
(upper panel of Fig. \ref{fig:sfe}).
Such a rapid decrease
of the SFE as the SB evolves  suggests that feedback plays a major
role in the evolution of these  objects.
$E_{\rm{SB}}$ is a quantity  derived from models and  it would be
more desirable to put the above finding on observational grounds.
Vega et al.
(2005) found a tight correlation between $E_{\rm{SB}}$ and the FIR
colour $f_{60}/f_{100}$ in the sense that as the SB evolves, i.e.
$E_{\rm{SB}}$ increases, the $f_{60}/f_{100}$ decreases. Thus, by
using the FIR colour instead of $E_{\rm{SB}}$, we obtain the
correlation shown in the lower panel of Fig. \ref{fig:sfe},
\begin{equation}
\log(SFE)= (1.74 \pm 0.2) \log(\frac{f_{60}}{f_{100}})+(1.55 \pm 0.03)
\end{equation}
GS04 claim that the SFE is constant for all galaxies and that the
star formation law in terms of dense molecular content is well
represented by a  power law with exponent $\sim$1.0.  Instead of SFE
we could have used the ratio L$_{\rm{IR}}$/L$_{\rm{HCN}}$ and
obtained a correlation analogous to that shown in Figure 9a of
GS04. Considering only ULIRGs in that figure, a definite trend of
decreasing efficiency at cooler $f_{60}/f_{100}$ colour, with slope
$\sim 1.72$, is seen. Therefore, this  correlation between the
observed colour  and the observed SFE indicates that the variation
of the SFE is real. The additional information on the evolutive
phase of the SB and on the AGN contamination, allowed by a realistic
modelling procedure, eliminates a large fraction of the dispersion in
the SFE, found by GS04.

\footnotesize
 Acknowledgements. O.V. and M.C. acknowledge the
support of the I.N.A.F. research fellowships. O.V. acknowledges the
support of the INAOE. A.B., G.L.G. and L.S. acknowledge the warm
hospitality of INAOE and the partial funding by the European
Community by means of the Maria Curie contract MRTN-CT- 2004-503929,
MAGPOP. Financial contribution from contract ASI-COFIS I/016/07/0
is  also acknowledged. This work is  based on data taken with the
\emph{Spitzer} Space Telescope, which is operated by the JPL,Caltech
under a contract with NASA.


\normalsize

\begin{thebibliography}
\bibitem[Armus et al.(2007)]{2007ApJ...656..148A} Armus, L., et al.\ 2007,
\apj, 656, 148
\bibitem[Blain et al.(2002)]{2002PhR...369..111B} Blain, A.~W., Smail, I.,
Ivison, R.~J., Kneib, J.-P., \& Frayer, D.~T.\ 2002, \physrep, 369,
111
\bibitem[Bressan et
al.(2002)]{2002A&A...392..377B} Bressan, A., Silva, L., \& Granato,
G.~L.\ 2002, \aap, 392, 377
\bibitem[Clemens et
al.(2008)]{2008A&A...477...95C} Clemens, M.~S., Vega, O., Bressan,
A., Granato, G.~L., Silva, L., \& Panuzzo, P.\ 2008, \aap, 477, 95
\bibitem[Condon et al.(1991)]{1991ApJ...378...65C} Condon, J.~J., Huang,
Z.-P., Yin, Q.~F., \& Thuan, T.~X.\ 1991, \apj, 378, 65

\bibitem[Farrah et al.(2001)]{2001MNRAS.326.1333F} Farrah, D., et al.\
2001, \mnras, 326, 1333
\bibitem[Farrah et al.(2003)]{2003MNRAS.343..585F} Farrah, D., Afonso, J.,
Efstathiou, A., Rowan-Robinson, M., Fox, M., \& Clements, D.\ 2003,
\mnras, 343, 585
\bibitem[Gao
\& Solomon(2004)]{2004ApJ...606..271G} Gao, Y., \& Solomon, P.~M.\
2004, \apj, 606, 271, (GS04)
\bibitem[Graci{\'a}-Carpio et al.(2006)]{2006ApJ...640L.135G}
Graci{\'a}-Carpio, J., Garc{\'{\i}}a-Burillo, S., Planesas, P., \&
Colina, L.\ 2006, \apjl, 640, L135
\bibitem[Granato
\& Danese(1994)]{1994MNRAS.268..235G} Granato, G.~L., \& Danese, L.\
1994, \mnras, 268, 235
\bibitem[Mihos
\& Hernquist(1996)]{1996ApJ...464..641M} Mihos, J.~C., \& Hernquist,
L.\ 1996, \apj, 464, 641
\bibitem[Mortier et al.(2005)]{2005MNRAS.363..563M} Mortier, A.~M.~J., et
al.\ 2005, \mnras, 363, 563
\bibitem[Nagar et
al.(2003)]{2003A&A...409..115N} Nagar, N.~M., Wilson, A.~S., Falcke,
H., Veilleux, S., \& Maiolino, R.\ 2003, \aap, 409, 115
\bibitem[Risaliti et al.(2006)]{2006MNRAS.365..303R} Risaliti, G., et al.\
2006, \mnras, 365, 303
\bibitem[Sanders et al.(1988)]{1988ApJ...325...74S} Sanders, D.~B. et al.\ 1988, \apj, 325, 74
\bibitem[Silva et al.(1998)]{1998ApJ...509..103S} Silva, L., Granato,
G.~L., Bressan, A., \& Danese, L.\ 1998, \apj, 509, 103
\bibitem[Smith et al.(1998)]{1998ApJ...493L..17S} Smith, H.~E., Lonsdale,
C.~J., Lonsdale, C.~J., \& Diamond, P.~J.\ 1998, \apjl, 493, L17
\bibitem[Vega et al.(2005)]{2005MNRAS.364.1286V} Vega, O., Silva, L.,
Panuzzo, P., Bressan, A., Granato, G.~L., \& Chavez, M.\ 2005,
\mnras, 364, 1286
\bibitem[Vega et
al.(2008)]{2008A&A...484..631V} Vega, O., Clemens, M.~S., Bressan,
A., Granato, G.~L., Silva, L., \& Panuzzo, P.\ 2008, \aap, 484, 631
\bibitem[Weedman et al.(2005)]{2005ApJ...633..706W} Weedman, D.~W., et al.\
2005, \apj, 633, 706



\end{thebibliography}
\end{document}